\documentclass[12pt]{article}
\def\be {\begin{equation}}
\def\ee {\end{equation}}
\def\ba {\begin{eqnarray}}
\def\ea {\end{eqnarray}}

\def\bi {\begin{itemize}}
\def\ei {\end{itemize}}
\begin{document}
\def\bea{\begin{eqnarray}}
\def\eea{\end{eqnarray}}

\title{\textbf{The Cardy-Verlinde formula and entropy of  the charged rotating BTZ black hole
}}

\author{  \textbf{M. R. Setare} \thanks{%
E-mail: rezakord@ipm.ir}\\{Department of Science, Payame Noor University, Bijar, Iran} \\
\textbf{ Mubasher Jamil} \thanks{%
E-mail: mjamil@camp.edu.pk
 }\\
{ Center for Advanced Mathematics and Physics, National University
of }\\{Sciences and Technology,  Rawalpindi, 46000, Pakistan}}

\maketitle

{\bf Abstract.} In this paper we show that the entropy of black hole
horizon in charged rotating BTZ space-time  can be described by the
Cardy-Verlinde formula, which is supposed to be an entropy formula
of conformal field theory in any dimension.

\section{Introduction}
The discovery of the existence of black hole solutions in three
spacetime dimensions by Ba\~nados, Teitelboin and Zanelli (BTZ)
\cite{Banados:1992wn,Banados:1992gq} ( for a review see Ref.
\cite{Carlip:1995qv}) represented one of the main recent  advances
for low-dimensional gravity theories. Owing to its simplicity and to
the fact that it can be formulated as a Chern-Simon theory, 3D
gravity as  become paradigmatic for understanding general features
of gravity, and in particular its relationship with gauge field
theories,  in any spacetime dimensions.

The realization of the existence of three dimensional (3D) black
holes not only deepened our understanding of 3D gravity but also
became a  central key for recent developments in gravity, gauge and
string theory.\\
The BTZ black hole continues to play a key role in  recent
investigations aiming to improve our understanding of 3D  gravity
and of  general feature of the gravitational interaction
\cite{Witten:2007kt}.

A characterizing feature of the  BTZ black hole (at least in its
uncharged form)  is the absence of  curvature singularities. The
scalar curvature is well-behaved (and constant) throughout the whole
3D spacetime. This feature is shared by other low-dimensional
examples such as 2D AdS black holes (see e.g. Ref.
\cite{Cadoni:1994uf}), for which also the microscopic entropy could
be calculated \cite{Cadoni:1998sg,Cadoni:1999ja} using the method
proposed in Ref. \cite{Strominger:1997eq}.

The absence of curvature singularities makes the BTZ black hole very
different from its higher dimensional cousins such as the 4D
Schwarzschild black hole. On the other hand one can  try to consider
low-dimensional black holes  with curvature singularities generated
by matter sources. In this paper we  consider the alternative case
in which the curvature singularity is not generated by mass sources
but by charges of the matter fields. An example, which we discuss in
this paper, is the electrically charged rotating BTZ (CR-BTZ) black hole.\\
One of the remarkable outcomes of the AdS/CFT correspondence has
been the generalization of Cardy's formula (Cardy-Verlinde formula)
for arbitrary dimensionality, as well as a variety AdS backgrounds.
The Cardy-Verlinde formula proposed by Verlinde \cite{ver}, relates
the entropy of a certain CFT with its total energy and its Casimir
energy in arbitrary dimensions. Quantum gravity in low-dimensional
anti-de Sitter(AdS) spacetime has  features that make it peculiar
with respect to the higher-dimensional cases. For $d=2,3$ the theory
is a conformal field theory (CFT) describing (Brown-Henneaux-like)
boundary   deformations and has a central charge determined
completely by Newton constant and the AdS length
\cite{Brown:1986nw,Cadoni:1998sg,
NavarroSalas:1999up,Maloney:2007ud}. Conversely, in  $d>4$, quantum
gravity in AdS spacetimes should admit a near-horizon description in
terms  of BPS solitons and D-brane excitations, whose low-energy
limit is an $U(N)$ gauge theory \cite{Strominger:1996sh,
Maldacena:1997re,Aharony:1999ti}. The difference between these two
descriptions is particularly evident in their application for
computing the entropy of non-perturbative gravitational
configurations such as black holes, black branes and
BPS states.\\
In the present paper we would like to check the consistency of the
Cardy-Verlinde formula, for the charged rotating BTZ black hole.
\section{The charged rotating BTZ black hole }
The BTZ black hole solutions \cite{Banados:1992wn,Banados:1992gq} in
$(2+1)$ spacetime dimensions are derived from a three dimensional
theory of gravity \be I=\frac{1}{16 \pi G}\int dx^{3} \sqrt{-g}\,(
R+2\Lambda)\label{ac1} \ee where $G$ is the three dimensional Newton
constant and $\Lambda=\frac{1}{l^2}>0$ is the cosmological constant.
Often in the literature units are chosen such that $G$ is
dimensionless, $8G=1$, here we use such units..
\par\noindent
The corresponding line element in Schwarzschild coordinates is \be
ds^2 =- f(r)dt^2 + f^{-1}dr^{2}¥+r^2\left(d\theta
-\frac{J}{2r^2}dt\right)^2 \label{metric}\ee whit metric function:
\be f(r)=\left(-M+\frac{r^2}{l^2} +\frac{¥J^2}{4
r^2}\right),\label{metric2}
 \ee where $M$ is the Arnowitt-Deser-Misner (ADM) mass,
$J$ the angular momentum (spin)
 of the BTZ black hole and $-\infty<t<+\infty$, $0\leq r<+\infty$,
 $0\leq \theta <2\pi$.
The outer and inner horizons, i.e. $r_{+}$ (henceforth simply black
hole horizon) and $r_{-}$ respectively, concerning the positive mass
black hole spectrum with spin ($J\neq 0$) of the line element
(\ref{metric}) are given as  \be
r^{2}_{\pm}=\frac{l^2}{2}\left(M\pm\sqrt{M^2 -
\displaystyle{\frac{J^2}{l^2}} }\right). \label{horizon1} \ee In
addition to the BTZ solutions described above, it was also shown in
\cite{Banados:1992wn,Martinez:1999qi} that charged black hole
solutions similar to (\ref{metric}) exist. These are solutions
following from the action \cite{Martinez:1999qi,Achucarro:1993fd}
\be I=\frac{1}{2\pi }\int dx^{3} \sqrt{-g}\,(
(R+2\Lambda-\frac{\pi}{2} F_{\mu\nu}F^{\mu\nu}) \label{ac2}. \ee The
Einstein equations are given by \be \label{ein}G_{\mu\nu}-\Lambda
g_{\mu\nu}=\pi T_{\mu\nu}, \ee where $T_{\mu\nu}$ is the
energy-momentum tensor of the electromagnetic field: \be \label{tmn}
T_{\mu\nu}=F_{\mu\rho}F_{\nu\sigma}g^{\rho\sigma}-\frac{1}{4}g_{\mu\nu}F^2.
\ee
 Electric charged black hole solutions of the equations (\ref{ein})
 takes the
form (\ref{metric}), but with \be \label{charged}
f(r)=-M+\frac{r^2}{l^2} +\frac{¥J^2}{4r^2}-\frac{\pi}{2} Q^2 \ln r,
\hspace{0.5cm} \ee whereas the $U(1)$ Maxwell field is given by
\be\label{maxw} F_{tr}=\frac{Q}{r},\label{metric3},
 \ee
where $Q$ is the
electric charge. Although these solution for
$r\to\infty$ are asymptotically AdS,
they have a power law curvature singularity at $r=0$,
 where $R\sim \frac{\pi Q^2}{r^2}$. This $r\to 0$ behavior of the
 Charged BTZ black hole has to be compared with that of the uncharged
 one, for which $r=0$ represent just a singularity of the causal
 structure . For $r> l$, the charged black hole is described by the
Penrose diagram as usual \cite{Kogan:1992nh}.

Horizons of the CR-BTZ metric are roots of the lapse function $f$.
Depending on these roots there are three cases of the CR-BTZ black
hole \cite{akbar} (see also\cite{kim}): Two distinct horizons
$r_{\pm}$ exist where plus correspond to the event horizon while
minus gives the Cauchy horizon (the usual CR-BTZ); black hole in
case of two repeated real roots gives a single horizon (extreme
case); and the case when no real root exists thus no horizon exists
(naked singularity).

We shall assume the first case in this paper. The black hole mass
and the angular momentum are given respectively by \be
M=\frac{r_+^2}{l^2}+\frac{J^2}{4r_+^2}-\frac{\pi}{2}Q^2\ln r \ee and
\be J=2r_+\sqrt{M-\frac{r_+^2}{l^2}+\frac{\pi}{2}Q^2\ln r_+} \ee
with the corresponding angular velocity to be
\be\Omega_+=\frac{J}{2r_+^2}=\frac{1}{r_+}\sqrt{M-\frac{r_+^2}{l^2}+\frac{\pi}{2}Q^2\ln
r_+}  \ee The Hawking temperature $T_{H}$ of the black hole horizon
is \be T_{H}=\left.\frac{df}{dr}\right\vert_{r=r_+}=\frac{1}{4\pi}
\Big(
\frac{2r_+}{l^2}-\frac{J^2}{2r_+^3}-\frac{\pi}{2}\frac{Q^2}{r_+}
\Big). \ee The entropy of the charged rotating BTZ black hole takes
the form \be \label{14} S_{BH}=4\pi r_+. \ee Also the electric
potential of the black hole is \be \Phi=\left.\frac{\partial
M}{\partial Q}\right\vert_{r=r_+}= -\pi Q \ln r_+.\ee The
generalized Cardy formula (hereafter named Cardy-Verlinde formula)
is given by \be \label{16} S_{SFT}=\frac{2\pi
R}{\sqrt{ab}}\sqrt{E_C(2E-E_C)}, \ee where $E$ is the total energy
and $E_C$ is the Casimir energy. The definition of the Casimir
energy is derived by the violation of the Euler relation \be
\label{17} E_C=n(E+PV-TS-\Phi Q-\Omega_+J), \ee where the pressure
of the CFT is defined as $P=E/nV$. The total energy may be written
as the sum of two terms \be \label{18} E=E_E+\frac{1}{2}E_C \ee
where $E_E$ is the purely extensive part of the total energy $E$.
The Casimir energy $E_C$ as well as the purely extensive part of
energy $E_E$ expressed in terms of the radius $R$ and the entropy
$S$ are written as \be \label{cas}E_C=\frac{b}{2\pi R} \ee \be
\label{ex} E_E= \frac{a}{4\pi R}S^2\ee
\section{Entropy of charged rotating BTZ black hole in Cardy-
Verlinde formula}
 The Casimir energy $E_C$, defined
as Eq.(\ref{17}), and $n=1$ in this case, is found to be \be
\label{21} E_C=\frac{1}{2}\Big( \frac{J^2}{r_+^2}+\pi Q^2 \Big) \ee
Additionally, it is obvious that \be \label{20} 2E-E_C=
\frac{2r^2_+}{l^2}- \pi Q^2\Big( \ln r_++\frac{1}{2} \Big) \ee The
purely extensive part of the total energy $E_E$ by substituting
Eq.(\ref{20}) in Eq.(\ref{18}), is given as\be \label{23} E_E=
\frac{r^2_+}{l^2} - \frac{\pi}{2} Q^2\Big( \ln r_++\frac{1}{2} \Big)
\ee whilst by substituting Eq.(\ref{14}) in Eq.(\ref{ex}), it takes
the form \be \label{24} E_E=\frac{4\pi a}{R}r_+^2 \ee Making use of
expression (\ref{cas}), Casimir energy $E_C$ can also be written as
\be \label{25} E_c=\frac{b}{2\pi R} \ee At this point it is useful
to evaluate the radius $R$. By equating the right hand sides of
(\ref{21}) and (\ref{25}), the radius is written as \be
R=\frac{b}{\pi \Big( \frac{J^2}{r_+^2}+\pi Q^2  \Big)}  \ee while by
equating the right hand sides (\ref{23}) and (\ref{24}) it can also
be written as \be R=\frac{4\pi ar_+^2l^2}{r_+^2-\frac{\pi}{2}
Q^2l^2\Big( \ln r_++\frac{1}{2} \Big)}  \ee Therefore, the radius
expressed in terms of the arbitrary positive coefficients $a$ and
$b$ is \be \label{28} R=\frac{2r_+l\sqrt{ab}}{\sqrt{ \Big(
\frac{J^2}{r_+^2}+\pi Q^2  \Big)(r_+^2-\frac{\pi}{2} Q^2l^2\Big( \ln
r_++\frac{1}{2} \Big))}} \ee Finally, we substitute expressions
(\ref{21}), (\ref{20}) and (\ref{28}) which were derived in the
context of thermodynamics of the charged rotating BTZ black hole, in
the Cardy- Verlinde formula (\ref{16}) which in turn was derived in
the context of CFT, and we get \be S_{CFT}=S_{BH}. \ee It has been
proven that the entropy of the charged rotating BTZ black hole can
be expressed in the form of Cardy-Verlinde formula.

\section{Conclusion}The Cardy-Verlinde formula proposed by Verlinde \cite{ver}, relates the entropy of
a certain CFT to its total energy and Casimir energy in arbitrary
dimensions, which is shown to hold for topological
Reissner-Nordstrom \cite{set1} and topological Kerr-Newman
\cite{set2} black holes  in de Sitter spaces, Taub-Bolt-$AdS_{4}$
\cite{tab}, Kerr-(A)dS \cite{kle}. There are many other relevant
papers on the subject \cite{set3}, \cite{set4}, \cite{a1}. Thus, one
may naively expect that the entropy of all CFTs that have an
AdS-dual description is given as the form (\ref{16}). However, AdS
black holes do not always satisfy the Cardy-Verlinde formula
\cite{gib}. For systems that admit 2D CFTs as duals, the Cardy
formula \cite{car} can be applied directly. This formula gives the
entropy of a CFT in terms of the central charge $c$ and the
eigenvalue of the Virasoro operator $l_0$. However, it should be
pointed out that this evaluation is possible as soon as one has
explicitly shown (e.g using the $AdS_d/CFT_{d-1}$ correspondence)
that the system under consideration is in correspondence with a 2D
CFT \cite{{Cadoni:1998sg},{Cadoni:1999ja}}. The aim of this paper is
to further investigate the AdS/CFT correspondence in terms of
Cardy-Verlinde entropy formula. In this paper, we have shown that
the entropy of the black hole horizon of charged rotating BTZ
spacetime can also be rewritten in the form of Cardy-Verlinde
formula.

\end{document}